# Cost Models for Selecting Materialized Views in Public Clouds

**Romain Perriot**
*Clermont Université, CNRS, Université Blaise Pascal, LIMOS UMR 6158*
*romain.perriot@univ-bpclermont.fr*

**Jérémy Pfeifer**
*Clermont Université, CNRS, Université Blaise Pascal, LIMOS UMR 6158*
*jeremy.pfeifer@univ-bpclermont.fr*

**Laurent d'Orazio**
*Clermont Université, CNRS, Université Blaise Pascal, LIMOS UMR 6158*
*laurent.dorazio@univ-bpclermont.fr*

**Bruno Bachelet**
*Clermont Université, CNRS, Université Blaise Pascal, LIMOS UMR 6158*
*bruno.bachelet@univ-bpclermont.fr*

**Sandro Bimonte**
*IRSTEA, UR TSCF, Clermont-Ferrand*
*sandro.bimonte@irstea.fr*

**Jérôme Darmont**
*Université de Lyon (Laboratoire ERIC)*
*jerome.darmont@univ-lyon2.fr*

## Abstract

*Data warehouse performance is usually achieved through physical data structures such as indexes or materialized views. In this context, cost models can help select a relevant set of such performance optimization structures. Nevertheless, selection becomes more complex in the cloud. The criterion to optimize is indeed at least two-dimensional, with monetary cost balancing overall query response time. This paper introduces new cost models that fit into the pay-as-you-go paradigm of cloud computing. Based on these cost models, an optimization problem is defined to discover, among candidate views, those to be materialized to minimize both the overall cost of using and maintaining the database in a public cloud and the total response time of a given query workload. We experimentally show that maintaining materialized views is always advantageous, both in terms of performance and cost.*

## 1. Introduction

Recently, cloud computing, led by companies such as Google, Microsoft and Amazon, attracted special attention. This paradigm allows access to on-demand, configurable resources that can be quickly made available with minimal maintenance. According to the pay-as-you-go pricing model, customers only pay for resources (storage and computing) they actually use. Performance in the cloud usually relies upon the use of a large number of instances, with parallel computing being transparent to the user.

Data warehouses and OLAP (On-Line Analytical Processing) are technologies for decision support enabling the online analysis of large data volumes. These technologies rely on optimization techniques such as indexes, caches or denormalized logical models that allow multidimensional analysis (aggregations on multiple axes of analysis) while ensuring good performance.With the broader and broaderavailabilityof clouds, organizations tend to deploy data analytics in the cloud to benefit from computing power and cheap storage, and to eliminate maintenance costs.

In this article, we focus on issues related to materializingviews in the cloud and its impact on the pay-as-you-go pricing model. Materialized views are used to physically store the results of relevant and frequent queriesto reduce response time. A major challenge is to select the best views to materialize. Traditionally, the criteria used for view selection mainly include storage and maintenance costs(Aouiche & Darmont, 2009; Baril & Bellahsene, 2003). In the cloud, storage is virtually infinite, so storing all views could be envisaged. However, materialized viewsstill incur storage and maintenance costs. The performance optimization problem is then to find a trade-offbetween response time and costs, and depends on the needs and assets of a particular user. At one end of the spectrum, users under a hard budget constraint can accept long response times, while at the other end, users may disregard costs if they need very fast response.

We address the multi-criteria optimization problem of selecting a set of views to materialize in order to optimize both the budgetary cost of storing and querying a data warehouse in the cloud, and the overall response time. To achieve this goal, our main contribution is the design of cost models for storing, maintaining and querying materialized views in the cloud. This article extends our previous proposal(Nguyen, Bimonte, d'Orazio, & Darmont, 2012) in three ways. First, we proposemore flexible cost models that can be applied to different vendors. Second, we introduce a new formulation that solves the optimization problem using a CPLEX solver. Finally, our solution is experimentally validated with the Star Schema Benchmark(O'Neil, O'Neil, Chen, & Revilak, 2009).

The remainder of this paper is organized as follows. In Section 2, we provide the background information that is used throughout the paper. In Sections 3 and 4, we define cost models for cloud data management and materializing views, respectively. In Section 5, we describe the optimization process that is based on these cost models. In Section 6, we present an experimental evaluation and the first performance analyses of our models. In Section 7, we discuss the state of the art and compare it to our approach. Finally, in Section 8, we conclude this paper and hint at future research directions.

## 2. Background

We present in this section the background information related to view materialization in the cloud. We first introduce a simple fictitious use case that serves as a running example throughout this paper. Then, we describe different pricing models in the cloud. Finally, we briefly recall the principle of view materialization.

### 2.1. Running example

To illustrate our work, we rely on a simulated dataset storing the sales of an international supply chain. Business users need to analyze the total profit per day, month, and year; and per administrative department, region, and country.

Our full dataset stores 10 years (2000-2010) of sale data. Its size is 500 GB. We run over this dataset a query workload Q that includes such queries as Q1= "sales per year and country", whose processing time is 0.2 hour. The size of Q's result is 10 GB. A typical materialized view we may consider to optimize overall response time is V1 = "sales per month and country", whose processing time is 0.1 hour. The whole set of selected materialized views is denoted V. V's size is 50 GB. Finally, the times to process Q with and without exploiting V are 40 hours and 50 hours, respectively.

### 2.2. Cloud pricing policies

Cloud Service Providers (CSPs) supply a pool of resources, such as hardware (CPU, storage, networks), development platforms, and services. There are many CSPs on the market, such as Amazon, Google, and Microsoft. Each CSP offers different services and pricing. This paper relies on limited, yet representative enough, models that include the main, commonly billed elements, i.e., CPU, storage, and bandwidth consumption[1]. These models are fully compliant with both relational (Amazon RDS, SQL Azure, Google Cloud SQL) and data intensive systems (MapReduce, Pig, SCOPE, Hive, Jaql), as for now, query response times are considered parameters of these models.

In order for the reader to have an overview of the pricing policies taken into account in the proposed models, we present in this section an example for both Microsoft Azure and Amazon Web Service (AWS). Even if the performance (response times and storage volume) differs from a system to another, identical values will be used in the example for clarity reasons. The objective of this work is indeed not to compare the different providers.

| Instance configuration | Price per hour |
|---|---|
| t1.micro | $0.02 |
| m1.small | $0.06 |
| m1.medium | $0.12 |
| m1.large | $0.24 |
| m1.xlarge | $0.48 |

*Table 1. EC2 computing prices*

| Instance configuration | Price per hour |
|---|---|
| Extra small | $0.02 |
| Small | $0.09 |
| Medium | $0.18 |
| Large | $0.36 |
| Extra large | $0.72 |

*Table 2. Azure computing prices*

Microsoft Azure(Microsoft, 2013) and Amazon Elastic Compute Cloud (EC2) (Amazon, 2013)provide computing resources. Different instance configurations can be rented (micro, extra small, small, large, extra large, etc.) at various prices, as illustrated in *Table 1* and*Table 2*. For example, the costs for a small instance (consisting in a 1.7 GB RAM, 1 EC2 Computing Unit, 160 GB of local storage under Linux for Amazon EC2; and a 1.75 GB RAM, 1 Computing Unit, 224 GB of local storage under Windows for Azure), are respectively $0.06 and $0.09 per hour for Amazon and Azure.

---

[1] Future works shall propose a generic framework to map to any CSP.

| Data volume | Price per month |
|---|---|
| *Input data* | |
| Any input data | Free |
| *Output data* | |
| First 5 GB | Free |
| Up to 10 TB | $0.12 per GB |
| Next 40 TB | $0.09 per GB |
| Next 100 TB | $0.07 per GB |
| Next 350 TB | $0.05 per GB |

*Table 3. Amazon and Microsoft bandwidth prices*

Bandwidth consumption is billed with respect to data volume (*Table 3*). Within Amazon and Azure models, input data transfers are free, whereas output data transfer cost varies with respect to data volume. Note that the same prices are applied by both providers.

| Price per month |
|---|
| $0.10 per GB |

*Table 4. Amazon EBS storage prices*

| Data volume | Price per month |
|---|---|
| First 1 TB | $0.095 per GB |
| Next 49 TB | $0.08 per GB |
| Next 450 TB | $0.07 per GB |
| ... | |

*Table 5. Amazon S3 storage prices*

| Data volume | Price per month |
|---|---|
| First 1 TB | $0.053 per GB |
| Next 49 TB | $0.049 per GB |
| Next 450 TB | $0.045 per GB |
| ... | |

*Table 6. Microsoft Azure storage prices*

Finally, CSPs supply storage capabilities. Prices usually vary with respect to data volume. However, as mentioned previously, CSPs provide different services, the pricing model differing from one to another. Amazon EBS proposes a per instance model, whereas Amazon S3 and SQL Azure enable a global storage. Amazon EBS (*Table 4*) proposes a fixed price, whereas Amazon S3 (*Table 5*) and SQL Azure (*Table 6*) enable an earned rate when volume increases.

### 2.3. Materialized views

In Database Management Systems, a view is a virtual table associated to a query answer. Views help indirectly save complex queries, format the same data in different forms, support logical independence, and reinforce security by masking some pieces of data from unauthorized users. Materializing a view, i.e., storing it physically into a table, further helps improve response time by avoiding recomputing the corresponding query each time the view itself is queried. However, materialized views must be refreshed when source data are updated, which induces some maintenance overhead.

In this work, we assume that we have at our disposal a set of candidate views for materialization that have already been preselected by an existing view selection method (e.g., (Baril & Bellahsene, 2003)). We aim at choosing the best candidates with respect to the cloud's pay-as-you-go model, taking pricing constraints into account before any view materialization. Research perspectives include extending existing view selection algorithms to consider pricing aspects in order to supply a uniform process.

### 3. Cloud pricing models

This section presents general cost models for data management in the cloud, i.e., without considering the use of materialized views. In cloud computing, customers rent resources to a CSP to run some applications. Figure 1 recalls the costs involved (Section 2.2), i.e., bandwidth consumption for input data transfers and query result retrieval, data storage, and application processing time.

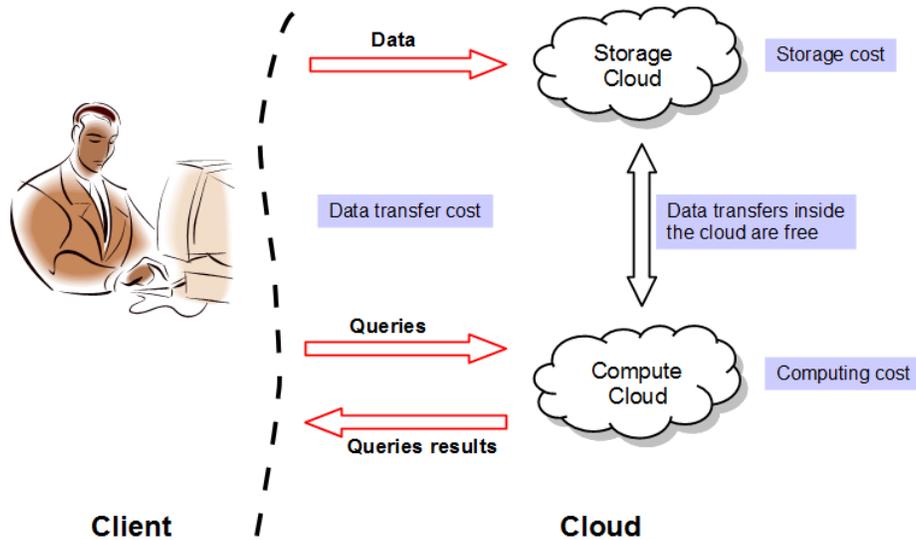

*Figure 1. Costs involved in cloud data management.*

Let $C_c$ be the sum of computing costs, $C_s$ be the sum of storage costs, and $C_t$ be the sum of data transfer costs. Then, the total cost $C$ for cloud data management is:

$$C = C_c + C_s + C_t. \qquad (1)$$

Let us define the general parameters and functions that we use to express our cost models (Table 7). Let $Q = \{Q_i\}_{i=1..n_Q}$ be the query workload and $A = \{A_i\}_{i=1..n_Q}$ the answers to the queries. The whole dataset is denoted D. Function $s(X)$ returns the size in GB of X, e.g., $s(A_i)$ is the size of the answer $A_i$. Function $t(X)$ returns the storage time of X, e.g., $t(D)$ is the storage time of dataset D in the cloud.

| Parameter | Description |
|---|---|
| $Q = \{Q_i\}_{i=1..n_Q}$ | Query workload |
| $A = \{A_i\}_{i=1..n_Q}$ | Query answers |
| $D = \{D_k\}_{k=1..n_D}$ | Dataset |
| $P$ | Cloud service provider |
| $IC = \{IC_j\}_{j=1..n_{IC}}$ | Instance configuration |
| $s(X)$ | Size in GB of $X$ |
| $t(X)$ | Storage time of $X$ |

*Table 7. General parameters.*

### 3.1. Piecewise linear functions

Some costs (transfer and storage costs especially) are piecewise linear functions. In this paper, we define a piecewise linear cost function $C(x)$ as a function decomposed into segments (Figure 2).

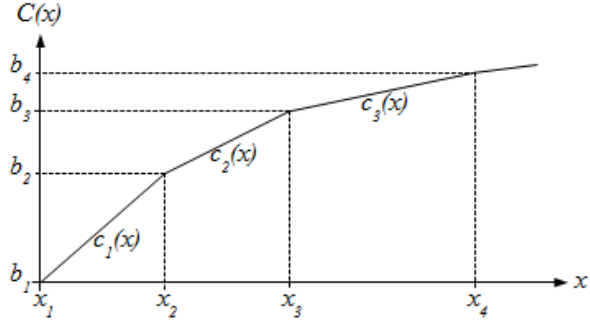

*Figure 2. Piecewise linear cost function.*

Each segment $e$ represents $C(x)$ in an interval of input values $[x_e; x_{e+1}[$ and is characterized by a gradient $a_e$ and an initial cost $b_e = C(x_e)$. Considering a segment $e$ such that $x \in [x_e; x_{e+1}[$, the cost $C(x) = c_e(x) = a_e \times (x - x_e) + b_e$. The piecewise linear function $C(x)$ is thus expressed as follows:

$$\begin{aligned} C(x) &= c_e(x) \\ &= a_e \times (x - x_e) + b_e \end{aligned} \qquad (2)$$

where $e$ is such that $x_e \leq x < x_{e+1}$.

### 3.2. Data transfer cost

Data transfer cost $C_t$ depends on the size of uploaded data, i.e., queries $Q_i$ and dataset $D$ (including the initial dataset and additional inserted data if any), on the size of downloaded data, i.e., query answers $A_i$, and on the pricing model applied by the provider P. The cost can be decomposed into an upload transfer cost $C_t^-$ and a download transfer cost $C_t^+$:

$$C_t(D, Q, A, P) = C_t^-(D, Q, P) + C_t^+(A, P). \qquad (3)$$

Note that most cloud providers, such as Amazon or Microsoft, do not charge for input data transfers, so input queries, initial and inserted data can be ignored for now. As a consequence, data transfer cost can be reduced to $C_t^+(A, P)$ and thus depends only on A and P:

$$C_t(A, P) = C_t^+(A, P). \tag{4}$$

The pricing of Azure and Amazon EC2 is variable. It is null for the first 5 GB. Then, it becomes $0.12 per GB up to 10 TB, $0.09 per GB up to 40 TB and so on (see Section 2.2). The cost function $C_t$ of such a pricing policy is a piecewise linear function (cf. Formula 2).

**Example 1.** *In our running example, with 10 GB of bandwidth consumption, $x_2^P \leq s(A) < x_3^P$, so data transfer cost is:* $C_t(A, P) = c_2^P(s(A)) = 0.12 \times (10 - 5) + 0 = \$0.60$.

### 3.3. Computing cost

Computing cost $C_c$ depends on the workload Q, the instance configuration IC, i.e., the type (micro, small, medium, etc.) and the number of nodes to be used, and the pricing model applied by provider P: $C_c(Q, IC, P)$.

Both Amazon and Microsoft associate a price with a type of instance. Each instance may bear variable performances (with respect to its number of CPUs, its available RAM, etc.), and thus different costs. Providers then compute the price to be paid by instance as the product of usage time by instance price. Finally, they sum the results for all allocated instances.

Let us set that queries are executed on an instance configuration IC composed of $n_{IC}$ computing instances $IC_j$: $IC = \{IC_j\}_{j=1..n_{IC}}$. The cost for renting instance $IC_j$ is denoted $c_c^P(IC_j)$. Processing time of query $Q_i$ on instance $IC_j$ is denoted $t^P(Q_i, IC_j)$. Then, the processing cost of running the set of queries $Q = \{Q_i\}_{i=1..n_Q}$ can be expressed by the following function:

$$C_c(Q, IC, P) = \sum_{i=1}^{n_Q} \sum_{j=1}^{n_{IC}} t^P(Q_i, IC_j) \times c_c^P(IC_j). \tag{5}$$

**Example 2.** *In our running example, let us consider that query workload $Q = \{Q_1\}$ is processed in 50 hours on two small instances of Amazon EC2. Then, its processing cost is:* $C_c(Q, IC, EC2) = t^{EC2}(Q_1, IC_1) \times c_c^{EC2}(IC_1) + t^{EC2}(Q_1, IC_2) \times c_c^{EC2}(IC_2) = 50 \times 0.06 + 50 \times 0.06 = \$6$.

### 3.4. Storage cost

Storage cost $C_s$ depends on the size and storage time of the whole dataset D, the instance configuration IC, and the pricing policy of provider P: $C_s(D, IC, P)$.

Storage time t(D) of dataset D can be divided into periods such that $t(D) = \sum_{k=1}^{n_D} t(D_k)$, where $D_k$ represents the whole dataset for period k. In each period k, the size $s(D_k)$ of the stored data is fixed. Total storage cost is thus the sum of the price to be paid for each period:

$$C_s(D, IC, P) = \sum_{k=1}^{n_D} C_s(D_k, IC, P). \tag{6}$$

Within Amazon EBS, total storage cost is the product of data size by storage time, fixed price $c_s^{EBS}$ per GB, and the number of computing instances $n_{IC}$. Indeed, as mentioned previously, each EC2 instance depends on a different EBS volume. Then storage cost can be expressed by:

$$C_s(D, IC, EBS) = \sum_{k=1}^{n_D} c_s^{EBS} \times s(D_k) \times t(D_k) \times n_{IC}. \tag{7}$$

**Example 3.** *We use Amazon EBS for storage pricing (Table 4), two small EC2 instances and we consider that 0.5 TB (512 GB) data have been stored for 12 months. At the beginning of the 8$^{th}$ month, we insert 2 TB (2048 GB) of new data in the cloud. Thus we have two periods. The storage cost is: $C_s(D, IC, EBS) = c_s^{EBS} \times s(D_1) \times nb_{IC} \times t(D_1) + c_s^{EBS} \times s(D_2) \times nb_{IC} \times t(D_2) = 0.10 \times 512 \times 2 \times 7 + 0.10 \times (512 + 2048) \times 2 \times (12 - 7) = \$3276.8$.*

The pricing of Amazon S3 is variable. It is $0.095 per GB for the first TB. Then, it becomes $0.08 per GB up to 450 TB and so on (Section 2.2). Unlike EBS, S3's price is independent from the number of EC2 instances. This pricing $c_s^{S3}$ is a piecewise linear function (cf. Formula 2), and the storage cost can be expressed as follows:

$$C_s(D, IC, S3) = \sum_{k=1}^{n_D} c_s^{S3}(s(D_k)) \times t(D_k). \tag{8}$$

**Example 4.** *We use Amazon S3 for storage pricing (Table 5) and consider the same scenario as in Example 3. Thus we have two periods, and due to their data volume, segment 1 of the cost function is considered for period 1, and segment 2 for period 2. The whole storage cost is: $C_s(D, IC, S3) = c_1^{S3}(s(D_1)) \times t(D_1) + c_2^{S3}(s(D_2)) \times t(D_2) = (a_1^{S3} \times (s(D_1) - x_1^{S3}) + b_1^{S3}) \times t(D_1) + (a_2^{S3} \times (s(D_2) - x_2^{S3}) + b_2^{S3}) \times t(D_2) = (0.095 \times (512 - 0) + 0) \times 7 + (0.08 \times ((512 + 2048) - 1024) + 1024 \times 0.095) \times 5 = \$1441.28$.*

## 4. Cost models for materializing views in the cloud

This section presents cost models for materializing views in the cloud, relying on the cost models developed in Section 3. We assume here that queries are executed on a constant number $n_{IC}$ of identical instances $IC_0$ ($IC_j = IC_0, \forall j = 1..n_{IC}$). In future work, we shall consider the evaluation process on multiple, variable instances.

Let $V_{cand} = \{V_k\}_{k=1..n_V}$ be a set of candidate views for materialization provided by an existing view selection technique (cf. Section 2.3). In this section, we assume that views to be materialized have been selected from the client's side, outputting a final set of views $V \subset V_{cand}$ that are materialized in the cloud. The problem of choosing the best set of views $V$ from $V_{cand}$ based on the cost models presented here is addressed in the next section.

### 4.1. Data transfer cost

Materializing views helps save bandwidth and benefits from the computing performance of the cloud. With materialized views created in the cloud, transfer costs due to materialization are null. As a consequence, total transfer cost $C_t$ is not impacted and remains expressed by Formula 4.

### 4.2. Computing cost

Using materialized views implies modifying the computing cost model, since query processing may exploit materialized views, and views must be materialized and maintained. Computing cost $C_c$ now depends on V: $C_c(Q, V, IC, P)$.

Applying Amazon and Microsoft pricing model with materialized views and considering that the cloud consists of a constant number of identical instances[2], Formula 5 thus becomes:

$$C_c(Q,V,IC,P) = T(Q,V,P) \times c_c^P(IC_0) \times n_{IC}. \tag{9}$$

$T(Q, V, P)$ is the total computing time, which is the sum of the time $T_{proc}(Q, V, P)$ for processing the queries in workload Q using the set V of materialized views, the time $T_{mat}(V, P)$ for materializing these views, and the time $T_{maint}(V, P)$ for maintaining them. As a consequence, the total computing time can be expressed as:

$$T(Q,V,P) = T_{proc}(Q,V,P) + T_{mat}(V,P) + T_{maint}(V,P). \tag{10}$$

If a view is materialized, its associated query must be executed, which must be paid for in the cloud. Let the materialization time of view $V_k$ be $t_{mat}^P(V_k)$. The total materialization time is:

$$T_{mat}(V,P) = \sum_{V_k \in V} t_{mat}^P(V_k). \tag{11}$$

The maintenance cost of materialized views is directly proportional to the time required for updating materialized views when they are impacted by modifications of the source dataset. Note that we consider that querying and maintenance do not occur at the same time. For example, queries are posed during day-time and maintenance is performed during night-time. Let the maintenance time of view $V_k$ be $t_{maint}^P(V_k)$. Then, the total maintenance time of V is:

$$T_{maint}(V,P) = \sum_{V_k \in V} t_{maint}^P(V_k). \tag{12}$$

When using materialized views, query processing time is defined by two main parameters: query workload Q and set of materialized views V. Queries may use the contents of materialized views instead of recomputing their result. Note that we consider that Q is fixed, variable workload is left for future work. Since views are usually materialized with respect to a given workload and ours is fixed, then V is also fixed. Let $t^P(Q_i, V)$ be the processing time of query $Q_i$ when exploiting the set of materialized views V. Thus, the total processing time of Q against V is:

$$T_{proc}(Q,V,P) = \sum_{i=1}^{n_Q} t^P(Q_i, V). \tag{13}$$

---

[2]Considering variable instances is out of the scope of thispaper and is part of our perspectives.

### 4.3. Storage cost

Using materialized views does not impact the storage cost model as presented by Formulas 6, 7 and 8. Exploiting materialized views to enhance query performance implies storing them in the cloud and paying the corresponding cost. As a consequence, some data can be duplicated. In that case, the size of $D + V$, i.e., $s(D) + \sum_{V_k \in V} s(V_k)$, is used instead of the size $s(D)$ of $D$ alone in the storage cost model. Therefore, the storage cost model depending on $D$ and $V$ can be expressed as:

$$C_s(D, V, IC, P) = C_s(D + V, IC, P). \tag{14}$$

Note that we assume that original data and materialized views are stored for the whole considered storage period.

**Example 5.** *In our running example with Amazon S3, the dataset (0.5 TB) has been stored for a year, the size of duplicated data due to materialized views is 50 GB. In addition, no data are inserted during the considered period. Thus, we have a single period, and storage cost is $C_s(D, V, IC, S3) = (512 + 50) \times 0.095 \times 12 = \$640.68$.*

## 5. Optimizing view materialization in the cloud

In this section, we investigate how to select the views to materialize in order to improve query performance with a minimum overhead of storage cost. We define optimization problems to select the best set of materialized views by exploiting the cost models introduced in Section 4. These problems are expressed here as linear programs with continuous and integer variables, in order to be solved efficiently using a mixed-integer programming (MIP) solver such as CPLEX[3]. However, for large instances, MIP solvers could not provide an optimal solution in a limited time. Thus, a GRASP heuristic (Feo & Resende, 1995) is proposed to find good solutions faster.

### 5.1. Optimization objectives

Based on the ideas in(Kllapi, Sitaridi, Tsangaris, & Ioannidis, 2011), we propose three optimization problems, labelled $MV_1$ to $MV_3$, with different objective functions to satisfy the needs and capacity of customers.

**Problem $MV_1$**: find a set of views $V$ that minimizes response time $T_{proc}$ under budget limit $C_{max}$ for total cost $C$.

$$(MV_1) \begin{cases} \text{minimize} & T_{proc} \\ \text{subject to} & C \leq C_{max} \\ & and\ constraints\ that\ define: \\ & -\ the\ cost\ model \\ & -\ the\ selection\ of\ views \end{cases} \tag{15}$$

**Problem $MV_2$**: find a set of views $V$ that minimizes total cost $C$ under limit $T_{max}$ for response time $T_{proc}$.

---
[3]IBM ILOG CPLEX Optimizer: http://www.ibm.com/software/integration/optimization/cplex-optimizer

$$(MV_2) \begin{cases} \text{minimize} & C \\ \text{subject to} & T_{proc} \leq T_{max} \\ & \text{and constraints that define:} \\ & \quad - \text{the cost model} \\ & \quad - \text{the selection of views} \end{cases} \quad (16)$$

Solving the bi-objective optimization problem (i.e., to minimize both C and $T_{proc}$) is not directly addressed in this paper, as it requires specific optimization techniques (e.g.,(Coello, Lamont, & Veldhuisen, 2007)) that differ from mono-objective optimization techniques. It is not possible to provide a solution that optimizes both objectives. However, non-dominated solutions can be provided (i.e., solutions of the Pareto frontier), meaning solutions that can not be improved on one objective without worsening the other objective (Ehrgott, 2005). The model presented here is fully valid for multi-objective optimization, by removing budget and response time limits.

We can consider using exact methods (e.g., Two-Phases Method (Visée, Teghem, Pirlot, & Ulungu, 1998)) or approximate methods (e.g., NSGA-II, Non-dominated Sorting Genetic Algorithm-II(Deb, Pratap, Agarwal, & Meyarivan, 2002)) to solve the bi-objective problem. The former approach is based on solving the following problem $MV_3$, which aims at optimizing both objectives with a coefficient α setting the relative importance of the response time criterion against the cost criterion.

**Problem $MV_3$**: find a set of views V that is a trade-off between minimum response time $T_{proc}$ and minimum total cost C.

$$(MV_3) \begin{cases} \text{minimize} & \alpha \times T_{proc} + (1-\alpha) \times C \\ \text{subject to} & \text{constraints that define:} \\ & \quad - \text{the cost model} \\ & \quad - \text{the selection of views} \end{cases} \quad (17)$$

**5.2. Mixed integer programming formulation**

In order to select the set of views V to materialize, we rely on an existing algorithm, such as(Baril & Bellahsene, 2003), enabling to obtain a set of candidate views for materialization $V_{cand} = \{V_k\}_{k=1..n_V}$. In a first step, we make some assumptions on $V_{cand}$. Notably, to determine the gain of using a view, or several views, for a given query requires either to know enough details on the functioning of the cloud to be able to express analytically this gain, or to run many experiments to measure or estimate the gain.

To simplify matters, we assume here that each query $Q_i$ can use only one single view among a set of candidate views, meaning that for each query $Q_i$, there is a set $V^i \subset V$ of candidate views, and no more than one view in this set must be selected for query $Q_i$. In future work, one can consider the candidate views of query $Q_i$ to be $V^i = \{V^{ij}\}_{j=1..n_i}$ that contains $n_i$ candidate sets of views $V^{ij} \subset V$. The objective will thus be to select a set of views $V^{ij}$ for a query $Q_i$ instead of a single view.

The decision of the optimization problem is to select a view $V_k$ for each query $Q_i$. For this purpose, decision variables $x_{ik}$ are introduced: $x_{ik} = 1$ if query $Q_i$ uses $V_k$, and $x_{ik} = 0$ otherwise. Let us define $g_{ik}^P$ as the gain on response time for query $Q_i$ when using view $V_k$ with cloud provider P. Gains are considered constant in this problem, meaning that they have been measured or estimated upstream (from experiments, statistics, or models). The response time of query $Q_i$ using views is expressed as follows:

$$t^P(Q_i, V) = t_i^P - \sum_{k=1}^{n_V} g_{ik}^P x_{ik}, \qquad (18)$$

where $t_i^P$ is the response time with cloud provider P for query $Q_i$ without any view, and assuming that no more than one view is selected for query $Q_i$. This latter point is ensured by the following constraints:

$$\sum_{k=1}^{n_V} x_{ik} \leq 1, \ \forall i = 1..n_Q. \qquad (19)$$

Decision variables $y_k$ are also introduced to determine whether view $V_k$ is materialized: $y_k = 1$ if view $V_k$ is materialized, and $y_k = 0$ otherwise. A view $V_k$ is materialized if it is used by at least one query (i.e., when at least one query $Q_i$ exists such that $x_{ik} = 1$), which is expressed by the following constraints:

$$x_{ik} \leq y_k, \ \forall i = 1..n_Q, \ \forall k = 1..n_V. \qquad (20)$$

Moreover, there is no need to materialize $V_k$ if it is not used at all, which is expressed by the following constraints:

$$y_k \leq \sum_{i=1}^{n_Q} x_{ik}, \ \forall k = 1..n_V. \qquad (21)$$

As for the gains on response time, materialization and maintenance times ($t_{mat}^P(V_k)$ and $t_{maint}^P(V_k)$) have been estimated upstream. These times must be considered for a view $V_k$ only if this view is materialized:

$$T_{mat}(V, P) = \sum_{k=1}^{n_V} t_{mat}^P(V_k) y_k,$$
$$T_{maint}(V, P) = \sum_{k=1}^{n_V} t_{maint}^P(V_k) y_k. \qquad (22)$$

As long as we assume that no data are inserted in the dataset during operation, we can consider a single period of length t(D) for storage. The size S of these data is:

$$S = s(D) + \sum_{k=1}^{n_V} s(V_k) y_k, \qquad (23)$$

and their storage cost is:

$$C_s(D, V, IC, P) = c_s^P(S) \, t(D) \, n_s^P, \qquad (24)$$

where $c_s^P$ is the storage cost function (we assume that it is piecewise linear, Section 3.4) applied by provider P, and $n_s^P = n_{IC}$ if provider P enables global storage (like Amazon EBS, Section 3.4) or $n_s^P = 1$ otherwise.

Note that transfer cost $C_t$ is not impacted by the selection of views. It is a constant value in the optimization problem and remains expressed by Formula 4. To sum up, the whole optimization problem, for instance $MV_1$ ($MV_2$ and $MV_3$ being very similar), is finally expressed as follows[4]:

$$MV_1 \begin{cases} \text{minimize} \quad T_{proc} = \sum_{i=1}^{n_Q} \left( t_i - \sum_{k=1}^{n_V} g_{ik}\, x_{ik} \right) \\[1em] \text{subject to} \quad C = C_c + C_t + C_s \leq C_{max} \qquad C_s = c_s(S)\, t(D)\, n_s \\[0.5em] \qquad C_c = (T_{proc} + T_{mat} + T_{maint})\, c_c(IC_0)\, n_{IC} \quad S = s(D) + \sum_{k=1}^{n_V} s(V_k)\, y_k \\[0.5em] \qquad \sum_{k=1}^{n_V} x_{ik} \leq 1,\ \forall i = 1..n_Q \qquad T_{mat} = \sum_{k=1}^{n_V} t_{mat}(V_k)\, y_k \\[0.5em] \qquad x_{ik} \leq y_k,\ \forall i = 1..n_Q,\ \forall k = 1..n_V \qquad T_{maint} = \sum_{k=1}^{n_V} t_{maint}(V_k)\, y_k \\[0.5em] \qquad y_k \leq \sum_{i=1}^{n_Q} x_{ik},\ \forall k = 1..n_V \\[0.5em] \qquad x_{ik} \in \{0,1\},\ \forall i = 1..n_Q,\ \forall k = 1..n_V \\[0.5em] \qquad y_k \in \{0,1\},\ \forall k = 1..n_V \end{cases} \quad (25)$$

Notice that $c_s$ is piecewise linear. It can be reformulated with linear constraints on continuous and integer variables (cf. (Chen et al., 2010), page 64), making the formulation fully linear.

### 5.3. GRASP heuristic

GRASP (Greedy Randomized Adaptive Search Procedure) is a metaheuristic with two phases: a randomized construction and a local search (Feo & Resende, 1995). The two phases are repeated a given number of times($it_{GR}$ times), and the best solution of all iterations is kept. In the construction phase, which is a greedy approach, a solution is iteratively constructed, by adding one element at a time in the solution. Then, the local search iteratively improves the solution obtained in the first phase by moving from solution to solution in the space of candidate solutions (by adding or removing elements in the solution).

There are two sets of decision variables in the view materialization problem: $y_k$ that indicates whether view $V_k$ is materialized, and $x_{ik}$ that indicates whether view $V_k$ is used by query $Q_i$. Note that if all $y_k$ are fixed, then finding the optimal values for all $x_{ik}$ is straightforward: selecting the materialized view $V_k$ that maximizes gain $g_{ik}$ for each query $Q_i$ provides an optimal solution for $x_{ik}$, since all $y_k$ are fixed.

In our GRASP heuristic, a solution is thus represented by vector $y = (y_k)_{k=1..n_V}$, and adding an element in the solution means materializing a view (i.e., for a given k, set $y_k = 1$).

---

[4]For clarityreasons, notation issimplified: parameters $Q$, $D$, $V$, $IC$, and $P$ are masked.

### 5.3.1. Randomized construction

The construction phase starts with a solution where no view is materialized. Iteratively, one view is selected and materialized in the solution. The aim here is to generate a feasible solution, meaning that the cost C of final solution y must be lowerthan $C_{max}$. Therefore, only views that reduce cost C are inserted in the solution. Several indicators are necessary to estimate the impact of materializing a view.

Let $c_k$ be the sum of the cost $\Delta C_c(V_k)$ of materializing view $V_k$ (including processing the materialization and the maintenance of the view), and the cost $\Delta C_s(V_k)$ of storing the view:

$$
\begin{aligned}
c_k &= \Delta C_c(V_k) + \Delta C_s(V_k), \\
\Delta C_c(V_k) &= (t_{mat}(V_k) + t_{maint}(V_k))c_c(IC_0)\, n_{IC}, \\
\Delta C_s(V_k) &= (c_s(S + s(V_k)) - c_s(S))t(D)\, n_s.
\end{aligned}
\qquad (26)
$$

Let $g_k$ be the gain on total response time $T_{proc}$ of materializing view $V_k$, which is the sum of all the gains induced by the materialization of view $V_k$ on each query $Q_i$ ($V_k$ provides a gain for query $Q_i$ only if $g_{ik} > g_{il}$, where $V_l$ is the current view selected for query $Q_i$):

$$
g_k = \sum_{i=1}^{n_Q} \max\left(0,\; g_{ik} - \sum_{l=1}^{n_V} g_{il}\, y_l\right). \qquad (27)
$$

Let $w_k$ be the benefit of materializing view $V_k$, which is the difference between the cost of the gain on response time and the cost of materializing the view:

$$
w_k = g_k\, c_c(IC_0)\, n_{IC} - c_k. \qquad (28)
$$

At each iteration of the greedy construction, views that are not materialized yet are ranked according to $w_k$. Only the views with $w_k > 0$ are considered, and among a given proportion($sel_{RC}$ %) of the bestcandidates (i.e., with highest $w_k$), one is chosen randomly. This selected view is materialized in solution y, and the procedure repeats, until there is no more candidate view to add in the solution.

At the end of the procedure, if cost $C > C_{max}$ for solution y, then a new attempt to build a feasible solution is performed. If, after a given number $it_{RC}$ of attempts, no feasible solution is found, then the heuristic stops with no feasible solution.

### 5.3.2. Local search

The goal of the local search is to improve the feasible solution obtained from the randomized construction, by reducing total processing time $T_{proc}$. The procedure moves from solution to solution by adding a view to be materialized at each iteration. For this purpose, the indicator $g_k$ of each view $V_k$ that is not materialized yet is computed. Neighborhood solutions of y will be solutions with one more materialized view $V_k$ such that $g_k > 0$, and that are still feasible, i.e., such that $C - w_k \leq C_{max}$. The heuristic moves to the solution that is randomly selected among a given proportion($sel_{LS}$ %) of the best solutions (i.e., with highest $g_k$) of the neighborhood.

Note that each time a new view is materialized, it can make some already materialized views useless, meaning that there can exist materialized views that are not used anymore by any query. To not materialize such views reduces total cost without increasing total processing time. Therefore, such views are detected and removed from each new solution of the local search. The procedure ends when no more view can be added to improve the solution.

## 6. Experiments

This section describes our experimental environment and the results we achieved. Experiments arerunat once in the cloud and on the client's side. Both the dataset and materialized views are stored on the cloud and queries are processed on the cloud. View selection algorithms are executed at the client's. The idea is to select views at the client's and materialize them on the cloud.

### 6.1. Environment

We run our experiments on a cluster composed by 20 virtual machines with 8 GB hard drives, 2 GB of RAM and 1 vCPU. The physical architecture corresponds to four 2.21 GHz processors with 12 cores and 96 GB of RAM. All machines run Hadoop (version 0.20.2) and Pig (version 0.9.1). Since client configuration has no effect on the following results, we do not detail it.

There are two main ways to generate workloads to test this kind of configuration. The first is using real traces. This approach usually helps provide good estimation of real usecases. However, a trace represents only a particular case and does not allow fully representing reality. Furthermore, if the main objective is to understand why a solution fits in a particular context, the use of one trace will be insufficient to highlight all operating mechanisms. The second approach consists in using a synthetic workload. Its main drawback lies in its artificial nature, but it allows comparing many configurations. In summary, if traces are available, using them help choose the model and calibrate it. Choosing the model is very important to provide a good representation of the target context. Since our goal is to illustrate the interest of our approach in datawarehouses, we use the Star Schema Benchmark (version 2.1.8.18). Queries are written in Pig Latin and executed by the Pig compiler asMapReduce tasks using a Hadoop cluster.

### 6.2. Querying

We have tested optimizations$MV_1$, $MV_2$and$MV_3$ described in Section 5.1 on a 5.5 GB database and Amazon S3 and EC2 prices. Measures have been performedwith variable parameters. The first parameter is experiment duration, from 1 to 24 months. The second parameter is workload frequency, i.e., the number of times the workload is executed during the considered period, from 1 to 5executions per week. The third parameter is the number of nodes used to process queries, from 5 to 20. A typical experiment fixes two parameters and varies the last. Fixed values used for experiment duration, workload frequency and node number are 12, 4 and 10, respectively.

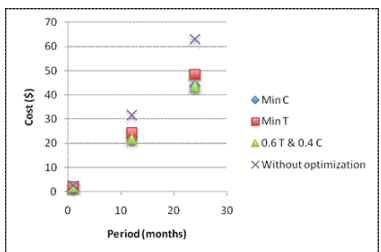 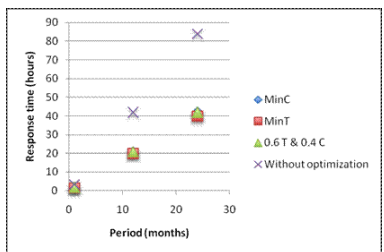 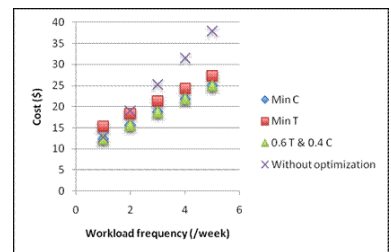

(a) Cost w.r.t. duration  (b) Response time w.r.t. duration  (c) Cost w.r.t. workload frequency

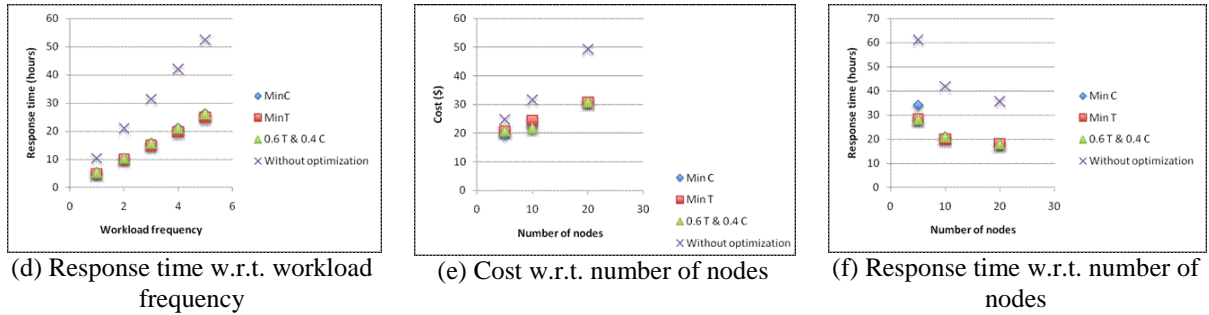

(d) Response time w.r.t. workload frequency  (e) Cost w.r.t. number of nodes  (f) Response time w.r.t. number of nodes

**Figure 3. Experimental results**

The experimental results we achieved clearly show that our approach allows selecting views that significantly improve both response time and cost (Figure 3). Response time can indeed be divided by about 2 (for example, with 4 workload executions per week during 12 months on a 10-node cluster, response time is 20 hours with materialized views, while it is 42 hours without), and cost is about 30% lower (for example, still with 4 workload executions per week during 12 months on a 10-node cluster, cost with materialized views is $22, while it is $32 without).

Our solution allows reaching the two objectives. When the objective is to reduce costs, materializing views allows paying the minimum for a given response. When the objective is performance, materializing views leads to the minimum response time for a given cost. However, note that both objectives do not seem contradictory. When fixing response time, we decrease cost by about 10%. When fixing cost, we improve response time by about 5%. But in large-scale environments, a gain of 5 to 10% is still estimated to thousands of dollars and hours of processing.

### 6.3. View selection

In this series of experiments, computational performance and solutions quality of the GRASP heuristic are compared with those obtained by solving the optimization problem $MV_1$ with CPLEX 12.4. The experiments are performed on an Intel Core 2 Quad Processor (2.5 GHz), with 4 GB of RAM. We empirically tested different parameter values for GRASP and retained the following: $it_{GR} = 100$, $it_{RC} = 200$, $sel_{RC} = 0.1$, and $sel_{LS} = 0.1$.

Solvers are tested on instances where values $s(V_k)$, $t_{mat}(V_k)$, $t_{maint}(V_k)$, $t_i$, $g_{ik}$ are randomly generated. Instances of various sizes (i.e., with different number $n_Q$ of queries, and number $n_V$ of candidate views) are built. The pricing of Amazon EC2 and S3 services is used, using 2 small computing instances for an operating period $t(D) = 1$.

Moreover, we test the same instances for three different values of $C_{max}$, in order to consider a more or less restrictive budget constraint. The minimum cost $C^-$ of the instance (obtained by solving problem $MV_2$ without response time limit) and the maximum cost $C^+$ of the instance (obtained by solving problem $MV_1$ without budget limit) are used to define three values for $C_{max}$: $C_1 = C^- + 0.05(C^+ - C^-)$ is 5 % above minimum cost $C^-$, $C_2$ is 15 % above $C^-$, and $C_3$ is 25 % above $C^-$. Therefore, three groups are formed, denoted $G_1$, $G_2$, and $G_3$, with $C_{max}$ being respectively equal to $C_1$, $C_2$, and $C_3$.

For each group, two tables are presented: on the left-hand table, the number of views is fixed to 100 and the number of queries varies. On the right-hand table, the number of queries is fixed to 100 and the number of views varies. Tables feature average values from 5 different instances with the same size. They show the CPU time needed by CPLEX and GRASP to solve the problem and the relative difference (gap) between the response time $T_{proc}$ of the solutions found by both methods. A gap of n% means that GRASP achieved a response time n% greater than that found by CPLEX. Note that if CPLEX cannot find the optimal solution within two minutes, the gap is computed with the best solution found by CPLEX (such cases are marked with ∗ in tables).

| Queries ($n_Q$) | CPU time (seconds) CPLEX | GRASP | Gap (%) |
|---|---|---|---|
| 10 | 0.1 | 0.0 | 0.5 |
| 20 | 2.6 | 0.1 | 1.1 |
| 30 | 8.6 | 0.1 | 0.4 |
| 40 | 17.8 | 0.1 | 0.7 |
| 50 | 14.1 | 0.1 | 0.5 |
| 60 | 33.3 | 0.2 | 1.0 |
| 70 | 51.8 ∗ | 0.2 | 0.3 |
| 80 | 67.6 ∗ | 0.2 | 0.9 |
| 90 | 48.6 | 0.3 | 0.6 |
| 100 | 84.6 ∗ | 0.3 | 0.4 |

Table 8. Results for $G_1$, with $n_V = 100$.

| Views ($n_V$) | CPU time (seconds) CPLEX | GRASP | Gap (%) |
|---|---|---|---|
| 10 | 1.1 | 0.0 | 0.6 |
| 20 | 3.4 | 0.0 | 0.6 |
| 30 | 9.7 | 0.1 | 0.8 |
| 40 | 17.4 | 0.1 | 0.4 |
| 50 | 26.4 | 0.1 | 0.5 |
| 60 | 47.7 | 0.2 | 0.4 |
| 70 | 43.3 | 0.2 | 0.4 |
| 80 | 69.2 ∗ | 0.2 | 0.3 |
| 90 | 73.4 ∗ | 0.3 | 0.8 |
| 100 | 84.6 ∗ | 0.3 | 0.4 |

Table 9. Results for $G_1$, with $n_Q = 100$.

The results for group $G_1$ are presented in Tables 8 and 9. The time needed by CPLEX to solve the instances increases significantly with their size. For some instances, CPLEX cannot find an optimal solution within two minutes. As GRASP runs fast, i.e., in less than one second, it has difficulties to find an optimal solution. However, the response time $T_{proc}$ of its solutions is usually less than 1% greater than that of the best solutions found by CPLEX.

| Queries ($n_Q$) | CPU time (seconds) CPLEX | GRASP | Gap (%) |
|---|---|---|---|
| 10 | 0.4 | 0.0 | 1.5 |
| 20 | 1.8 | 0.1 | 1.2 |
| 30 | 4.3 | 0.1 | 0.5 |
| 40 | 3.5 | 0.1 | 0.3 |
| 50 | 10.4 | 0.2 | 0.3 |
| 60 | 10.7 | 0.2 | 0.3 |
| 70 | 21.1 | 0.3 | 0.2 |
| 80 | 27.6 | 0.3 | 0.3 |
| 90 | 17.1 | 0.4 | 0.4 |
| 100 | 45.1 | 0.5 | 0.2 |

Table 10. Results for $G_2$, with $n_V = 100$.

| Views ($n_V$) | CPU time (seconds) CPLEX | GRASP | Gap (%) |
|---|---|---|---|
| 10 | 0.6 | 0.0 | 1.6 |
| 20 | 2.2 | 0.0 | 0.6 |
| 30 | 6.5 | 0.1 | 0.4 |
| 40 | 9.8 | 0.1 | 0.3 |
| 50 | 13.5 | 0.2 | 0.3 |
| 60 | 13.8 | 0.2 | 0.3 |
| 70 | 14.0 | 0.3 | 0.3 |
| 80 | 38.9 | 0.3 | 0.3 |
| 90 | 53.0 ∗ | 0.4 | 0.3 |
| 100 | 45.1 | 0.5 | 0.2 |

Table 11. Results for $G_2$, with $n_Q = 100$.

The results for group $G_2$ are presented in Tables 10 and 11. With the budget constraint relaxed, CPLEX seems to solve the instances more easily, abouttwice faster, while GRASP finds better solutions (the gap is a little smaller).

| Queries ($n_Q$) | CPU time (seconds) CPLEX | CPU time (seconds) GRASP | Gap (%) |
|---|---|---|---|
| 10 | 0.2 | 0.0 | 1.0 |
| 20 | 0.5 | 0.1 | 0.7 |
| 30 | 0.2 | 0.1 | 0.0 |
| 40 | 0.8 | 0.2 | 0.1 |
| 50 | 0.6 | 0.2 | 0.0 |
| 60 | 0.8 | 0.3 | 0.0 |
| 70 | 5.1 | 0.3 | 0.0 |
| 80 | 2.6 | 0.4 | 0.0 |
| 90 | 1.5 | 0.5 | 0.0 |
| 100 | 1.0 | 0.6 | 0.0 |

*Table 12. Results for $G_3$, with $n_V = 100$.*

| Views ($n_V$) | CPU time (seconds) CPLEX | CPU time (seconds) GRASP | Gap (%) |
|---|---|---|---|
| 10 | 0.6 | 0.0 | 0.8 |
| 20 | 0.9 | 0.1 | 0.5 |
| 30 | 3.3 | 0.1 | 0.2 |
| 40 | 6.5 | 0.2 | 0.3 |
| 50 | 5.8 | 0.2 | 0.1 |
| 60 | 3.5 | 0.3 | 0.1 |
| 70 | 2.6 | 0.3 | 0.0 |
| 80 | 1.6 | 0.4 | 0.0 |
| 90 | 2.3 | 0.5 | 0.0 |
| 100 | 1.0 | 0.6 | 0.0 |

*Table 13. Results for $G_3$, with $n_Q = 100$.*

Finally, the results for group $G_3$ are presented in Tables 12 and 13. With the budget limit significantly loosened, CPLEX has no difficulty to solve the instances, while GRASP does not always find an optimal solution (even if for the biggest instances, optimal solutions are found).

## 7. Related works

We discuss in this section previous research related to the main domains addressed in this paper, i.e., cloud data management, data access optimization through materialized views, and cost models for large-scale distributed systems.

Cloud data management brought about a lot of research and various operational systems. The most popular include so-called NoSQL systems, such as Amazon DynamoDB (DeCandia et al., 2007), or Cassandra (Lakshman & Malik, 2009), which scale up very efficiently but only proposes eventual consistency, in contrast to traditional ACID (Atomicity, Consistency, Isolation, Durability) guarantees. Full cloud relational systems enforcing ACID constraints are also available, e.g., Microsoft SQL Azure (Campbell, Kakivaya, & Ellis, 2010), Amazon RDS (Amazon, 2013) and Oracle Database Cloud Service (Oracle, 2013), but they currently operate on a smaller scale. Finally, there exist large-scale data analytics systems that are specifically tailored for the cloud, such as Pig (Gates et al., 2009) and Hive (Thusoo et al., 2010). The solution we propose in this paper is generic and can be applied within any of these systems.

In the cloud, performance is mainly managed by exploiting computing power elasticity. Nevertheless, well-known performance optimization techniques from the database domain, such as indexing, view materialization or caching, may be used to decrease the global monetary cost of querying data in the cloud. In this paper, we particularly focus onto view materialization. Numerous approaches help select (Agrawal, Silberstein, Cooper, Srivastava, & Ramakrishnan, 2009; Ceri & Widom, 1991; Luo, Naughton, Ellmann, & Watzke, 2003; Mami & Bellahsene, 2012; Yang, Karlapalem, & Li, 1997; Zhou, Larson, & Elmongui, 2007) materialized views, whether in transactional databases, in decision-support databases (i.e., data warehouses) or even on the Web.

In the view selection problem we address, all combinations of attributes in a database constitute a lattice of candidate materialized views. Cost models then help determine the materialized views that allow the best global performance improvement, usually under disk space constraints. Various optimization techniques are used to exploit these cost models, ranging from simple greedy algorithms (Vijay Kumar & Ghoshal, 2009) to simulated annealing or genetic algorithms (Bellatreche et al., 2006). Finally, to reduce the dimensionality of the input candidate view set, materialized views may also be pre-filtered with respect to the query workload, e.g., with data mining techniques such as frequent itemset mining or clustering (Aouiche & Darmont, 2009). Our own costs models aim at extending existing materialized view selection strategies by substituting classical space constraints by the pay-as-you-go economic model of the cloud, in order to achieve the best trade-off between storage cost (including the cost of storing materialized views) and computation cost.

Existing cost models for cloud computing address a variety of problems. For instance, Kllapi et al. (2011) worked on data stream scheduling with respect to monetary cost and processing time. On each cloud node, they slice the time into windows. Financial cost is then assumed to be the count of the time windows that have at least one operator running, multiplied by the cost of leasing the node. Kantere, Dash, Gratsias, and Ailamaki (2011) sought to amortize the cost of data structures such as indexes and materialized views to ensure the economic viability of the cloud service provider. With the help of stochastic models, they compute, among other indicators, the influence Inf($S$) of the cost of a new data structure $S$ on the economy of the cloud service provider. Then, the number $n$ of amortization payments for building $S$ is such that the gain probability of at most $n$ payments is equal to Inf($S$). Dash, Kantere, and Ailamaki (2009) also worked on this topic, but for automatically managing caches. They consider the cost of a query plan $PQ$ as the sum of the cost of executing $PQ$ and the amortized cost of any structure used by $PQ$. Then, cost models for cache queries, network queries, building and maintaining caching structures are detailed. Finally, Upadhyaya, Balazinska, and Suciu (2012) envisaged the problem of selecting and pricing optimizations in the cloud as a mechanism design problem, i.e., maximizing the expected value to users of exploiting a set $A$ of optimizations minus the actual cost of $A$, including its maintenance.

Other costs models target specific applications, especially in the field of astronomy. In this domain, simulation experiments showed that a good trade-off between storing optimization structures and computing power helps reduce the global cost of data processing in the cloud (more precisely, in Amazon's free solution) without reducing performances (Deelman, Singh, Livny, Berriman, & Good, 2008). The performance of three applications managing data streams, bearing various characteristics in terms of I/Os and memory and CPU consumption, have also been compared on Amazon EC2 and a high-performance cluster, to identify what applications achieved the best performances at the lowest cost (Berriman et al., 2010). Our work complements these existing cost models, but also differs from them in two ways. First, we introduce a new billing model that is generic enough to represent the billing models of all cloud service providers we are aware of. Second, our proposal rests on a detailed model of the optimization process that leads to materialized view selection. Finally, our materialized view selection approach is also independent from any particular target application.

## 8. Conclusion

We propose in this paper an approach to improve data management in the cloud using materialized views. Our main contributions are extended cost models for materializing views that take existing cloud pricing models into account. Our cost models are then exploited by an optimization process, which provides a compromise between the performance improvement due to materialization and budgetary constraints. Experiments on a private data center have highlighted the relevance of our approach.

This work opens many perspectives. First, we aim at extending our costmodels to overcome some limits (multiple and variable instances, for example). Then, we plan to integrate our models in existing view selection algorithms to avoid splitting the selection process into two phases, i.e., we aim at fusing the candidate view generation and view selection processes. In addition, it would be relevant to consider other optimization techniques, such as indexing or caching. It has indeed been shown that, jointly employed, indexes and materialized views benefit from each other (Aouiche & Darmont, 2009). Finally, we plan to validate our proposals on a larger scale and cloud-specific query workloads (with good and bad candidates for parallelism).

# Cost Models for Selecting Materialized Views in Public Clouds

R. Perriot, J. Pfeifer, L d'Orazio; B. Bachelet, S. Bimonte, J. Darmont

## Acknowledgment

This work was supported by the French National Research Agency. We acknowledge the contributions to this work of Thi Van Anh Nguyen, Vilmar Jefté Rodrigues de Sousa, Michael David de Souza Dutra, and members of the ERIC, LIMOS and IRSTEA. We also thank the anonymous referees for many helpful comments.

# Cost Models for Selecting Materialized Views in Public Clouds

R. Perriot, J. Pfeifer, L d'Orazio; B. Bachelet, S. Bimonte, J. Darmont